\documentclass[aps,pra,superscriptaddress,twocolumn]{revtex4-1}

\usepackage{graphicx}
\usepackage{amssymb,amsmath,amsfonts,bm}

\newcommand{\ui}{\underline{i}}
\newcommand{\uj}{\underline{j}}
\newcommand{\uz}{{\underline{0}}}
\newcommand{\uo}{{\underline{1}}}

\newcommand{\un}{{\underline{n}}}

\newcommand{\ket}[1]{|{#1}\rangle}
\newcommand{\bra}[1]{\langle{#1}|}

\begin{document}
\title{99\%-fidelity ballistic quantum-state transfer through long uniform channels}

\author{T.~J.~G.  Apollaro}
\affiliation{Dipartimento di Fisica, Universit\`a della Calabria,
             Via P.Bucci, 87036 Arcavacata di Rende (CS), Italy}

\affiliation{INFN—--Gruppo collegato di Cosenza,
             Via P.Bucci, 87036 Arcavacata di Rende (CS), Italy}

\author{L. Banchi}
\affiliation{ISI Foundation, Via Alassio 11/c,
             I-10126 Torino (TO), Italy}

\author{A. Cuccoli}
\affiliation{Dipartimento di Fisica, Universit\`a di Firenze,
             Via G. Sansone 1, I-50019 Sesto Fiorentino (FI), Italy}

\affiliation{INFN Sezione di Firenze, via G.Sansone 1,
             I-50019 Sesto Fiorentino (FI), Italy}

\author{R. Vaia}
\affiliation{Istituto dei Sistemi Complessi,
             Consiglio Nazionale delle Ricerche,
             via Madonna del Piano 10,
             I-50019 Sesto Fiorentino (FI), Italy}

\author{P. Verrucchi}
\affiliation{Istituto dei Sistemi Complessi,
             Consiglio Nazionale delle Ricerche,
             via Madonna del Piano 10,
             I-50019 Sesto Fiorentino (FI), Italy}
\affiliation{Dipartimento di Fisica, Universit\`a di Firenze,
             Via G. Sansone 1, I-50019 Sesto Fiorentino (FI), Italy}
\affiliation{INFN Sezione di Firenze, via G.Sansone 1,
             I-50019 Sesto Fiorentino (FI), Italy}
\date{\today}

\begin{abstract}
Quantum-state transfer with fidelity higher than $0.99$ can be
achieved in the ballistic regime of an arbitrarily long
one-dimensional chain with uniform nearest-neighbor interaction,
except for the two pairs of mirror symmetric extremal bonds, say $x$
(first and last) and $y$ (second and last-but-one). These have to be
roughly tuned to suitable values $x\,{\sim}\,2N^{-1/3}$ and
$y\,{\sim}\,2^{3/4}N^{-1/6}$, $N$ being the chain length. The general
framework can describe the end-to-end response in different models,
such as fermion or boson hopping models and $XX$ spin chains.
\end{abstract}


\maketitle

\section{Introduction}

Transferring quantum states between distant registers is one of the
basic tasks that a quantum computer based on qubits located on fixed
positions has to accomplish. The general scheme where a quantum
channel physically connects the sending and the receiving qubits
naturally emerges from such requirement and a variety of proposals
have been recently put forward for realizing quantum channels for
state transmission by different physical solutions: phonon modes for
trapped ions~\cite{CiracZ1995} or photon modes for superconducting
qubits ~\cite{BlaisHWGS2004,Yang2010}, molecular lattices for
vibrational excitons~\cite{Pouthier2012} or optical lattices for cold
atoms~\cite{BriegelCJCZ2000,ShersonWECBK2010,LapasarKKNO2011,%
GiampaoloI2010}, arrays of coupled quantum
dots~\cite{NikolopoulosPL2004,PetrosyanL2006,Paganelli2009,YanggBB2010},
and interacting $S=\frac12$ spins on a one-dimensional lattice,
usually referred to as spin chains. In particular, proposals based on
spin chains were first introduced in this context by
Bose~\cite{Bose2003,Bose2007} and have attracted much attention in
the last decade, due to both the possibility of exploiting their
natural dynamics for the transfer process, and the availability of
analytical results that allow for a detailed description of their
dynamics.

All of the above mentioned proposals are based on the idea that the
state of the sender-qubit be transferred at large distance via a
dynamical site-to-site hopping mechanism that we will here describe
in terms of a general {\em hopping model}. Once a 2-dimensional
Hilbert space, generated by $\ket{0}$ and $\ket{1}$, is assigned to
each site $i=1,...,N$ of a one-dimensional lattice, the hopping model
is defined by the following Hamiltonian:
\begin{equation}
 {\cal{H}} = \frac12 \sum_{i,j=1}^N A_{ij}~\,|\ui\rangle\langle\uj|~,
\label{e.H-hopping}
\end{equation}
where
$|\ui\rangle\equiv\ket{0}^{\otimes{i-1}}\ket{1}\ket{0}^{\otimes{N-i}}$
is a brief notation for single-excitation states, and the structure
of the hopping-amplitude matrix $\bm{A}\,{=}\,\{A_{ij}\}$ depends on
the properties of the specific model. Despite being responsible of a
possibly long-distance transfer process, such mechanism can be
generated by a short-range interaction. Notice that
Eq.~\eqref{e.H-hopping} describes free bosons as well as free
fermions and, as far as the latter are concerned, an exact mapping
exists between the Hamiltonian~\eqref{e.H-hopping} and that of the
$S\,{=}\,\frac12$ $XY$ chain, where the matrix elements $A_{ij}$
correspond to the exchange couplings. By the hopping model the basic
transmission mechanisms can be studied and characterized: this, in
turn, can also enlighten on the specific role of
nonlinearity~\cite{BayatB2010,BBBV2011,BBVB2011},
noise~\cite{ZwickASO2012} and
dissipation~\cite{RafieeSMM2011,Hu2011}.

The quality of the channel, as far as the transfer process is
concerned, essentially depends on the hopping amplitudes $A_{ij}$; in
particular, it has been proven that locally engineered hopping
amplitudes can lead to perfect state
transfer~\cite{Kay2010,KarbachS2005,ChristandlDEL2004,YungB2005,%
DiFrancoPK2008,ZwickASO2011,WangSR2011}. However, strategies based on
a very detailed design of the internal couplings appear very
demanding from a practical point of view~\cite{CappellaroVR2011}.

Having in mind that experimental setups generally require the
couplings to be as uniform as possible~(see, e.g.,
Ref.~\cite{RamanathanCVC2011}), different authors have proposed
alternative strategies for obtaining high-quality state-transfer
processes through mirror-symmetric channels with uniform bulk and
just a few extremal couplings allowed to be
varied~\cite{WojcikLKGGB2005,ApollaroP2006,CamposBR2007,BACVV2010,%
FeldmanKZ2010,ZwickO2011}. Some of these alternative strategies are
based on the idea of markedly weakening the coupling between the
channel and the sender/receiver qubits, a solution that has been
shown~\cite{WojcikLKGGB2005,CamposBR2007,YaoJGGZDL2011} to lead to a
very high-quality state transfer. However, this scheme yields large
transmission times and requires such a reduction of the extremal
bonds that a severe limitation on the actual length of a functioning
channel must be taken into account.

In general, the transmission quality of almost uniform channels is
expected to deteriorate as the length of the channel is
increased~\cite{Bose2003,Bose2007,FeldmanKZ2010} due to dispersion,
which is integral to uniformly distributed couplings. On the other
hand, modeling a scalable quantum-state transfer process whose
quality depends as little as possible on the physical length of the
channel is an essential issue to address, especially if solid-state
implementations and/or experimental analysis are in order.

Therefore, we find ourselves squeezed between the seemingly
incompatible requirements of avoiding too much a detailed design of
the physical channel and yet getting a reasonably long channel
characterized by a relatively convenient transfer time.

Our approach for solving this puzzle stems from the idea of
exploiting, in an almost uniform channel, the ballistic
state-transfer mechanism that allows perfect transmission of a
wavepacket~\cite{OsborneL2004,BACVV2010,Yadsan-ApplebyO2012}, in
virtue of a perfectly coherent, non dispersive, dynamics.
Perfect-transfer follows from the requirement that the normal modes
correspond to equally spaced eigenfrequencies (or, loosely speaking,
to a linear dispersion relation), so that a coherent mirrored
reconstruction of all normal components occurs~\cite{AronsteinS1997}.
We infer that only the modes involved in the initial configuration of
the overall system need satisfying the linearity
condition~\cite{BACVV2010} in order to get an effectively ballistic
dynamics: systems realizing this condition can be dubbed `optimal
state-transfer' systems.

This idea has been first proposed and
implemented~\cite{BACVV2010,BACVV2011} in a scheme where we could
only play with one parameter of the Hamiltonian, namely the value of
the extremal bond $x=A_{12}=A_{N-1,N}$.

We have learned that the emergence of an optimal value for $x$
follows from quite a complex interplay between two conflicting
effects, namely the deformation of the eigenvalue spacings and the
shrinking of the mode distribution, which are simultaneously driven
by the value of $x$. In order to further improve our results, we
understand that these effects must be handled independently, a goal
that can be accomplished by introducing just one more parameter in
the model, as shown in this paper. Notice that the introduction of a
second parameter in the model should not be thought of as a way of
moving towards the perfect-transfer scheme (which requires $N/2$
parameters) but rather as a practical answer to the effective need of
controlling two competing effects.

In Section~\ref{s.hoppingmodel} the state-transfer mechanism in the
hopping model Eq.~\eqref{e.H-hopping} is studied and the quantities
that characterize the efficiency of the quantum channel in terms of
the transition amplitudes are obtained. In Section~\ref{s.QUchannel}
we describe the setup for implementing the two-parameter
optimal-transfer scheme (with two adjacent modified bonds at the ends
of the chain) and derive implicit analytical expressions for the
frequencies and the matrix elements entering the transition
amplitude. The behavior of the latter is thoroughly discussed in
Section~\ref{s.amplitude}, where numerical results are reported
together with the analytical derivation of the large-$N$ limit of the
attainable optimal transmission fidelity. In Section~\ref{s.dynamics}
the dynamical behavior is shown to be ballistic, i.e., the
information is carried by a wavepacket traveling at constant speed
along the channel. Conclusions are drawn in Section~\ref{s.concl}.
Some details of the calculations are reported in
Appendix~\ref{a.spectral}.

\section{State transfer in the hopping model}
\label{s.hoppingmodel}

Let us consider the Hamiltonian~\eqref{e.H-hopping}. In the case of
nearest-neighbor interaction the matrix $\bm{A}$ is tridiagonal and,
for the chain to be connected, the off-diagonal elements cannot
vanish. Without loss of generality~\cite{Parlett1998} we assume
$\bm{A}$ to be real with $A_{i,i+1}\,{=}\,A_{i+1,i}\,{>}\,0$.
Introducing the orthogonal matrix $\bm{U}=\{U_{ni}\}$ that
diagonalizes the matrix $A$,
\begin{equation}
 \sum_{ij}U_{ni}A_{ij}U_{mj}=\lambda_n\,\delta_{nm}
            \equiv 2\,\omega_n\,\delta_{nm}~,
\label{e.Uni}
\end{equation}
and the one-excitation states
\begin{equation}
 \ket{\un}=\sum_iU_{ni}\ket{\ui}~,
\end{equation}
the Hamiltonian turns into the diagonal form
\begin{equation}
 {\cal{H}} = \sum_{n=1}^N \omega_n\ket{\un}\bra{\un}~.
\end{equation}

If $\bm{A}$ is mirror-symmetric, i.e., it commutes with the {\em
mirroring matrix} $\bm{J}=\{J_{ij}{=}\delta_{N+1-i,j}\}$, then also
$\bm{J}$ is diagonalized by $\bm{U}$,
\begin{equation}
 \sum_jJ_{ij}U_{nj}=U_{n,N+1-i} =j_n\,U_{ni}~,
\label{e.jn}
\end{equation}
and since $\bm{J}^2=1$ the eigenvalues $j_n$ are either $1$ or $-1$.
It is proven in Ref.~\cite{CantoniB1976} that if the eigenvalues
$\omega_n$ are cast in decreasing order then
$j_n=(-)^{n+1}=-e^{\imath\pi{n}}$; in other words, the $n^{\rm{th}}$
eigenvector of $\bm{A}$ is mirror-symmetric or -antisymmetric
according to whether $n$ is odd or even.

\subsection{Estimating state-transfer quality}

The purpose of state transfer is to start with a product state
$\ket{\underline\psi}=\ket{\psi}\ket{0}^{\otimes{N-1}}$, with a
generic state
\begin{equation}
 \ket{\psi} =\alpha\,\ket{0}+\beta\,\ket{1}
\end{equation}
of the first qubit, and let it evolve with ${\cal{H}}$ in such a way
that at a given time $t\simeq{N}$ the state of the last qubit is as
close as possible to $\ket{\psi}$. Formally, the evolved overall
state is
$\rho(t)=e^{-\imath{\cal{H}}t}\ket{\underline\psi}\bra{\underline\psi}
e^{\imath{\cal{H}}t}$ and the instantaneous state at site $i$ is
\begin{equation}
 \rho_i(t)={\textstyle{\rm{Tr}}_{\{1,...,i{-}1,i{+}1,...,N\}}}\rho(t)~.
\end{equation}
Defining $\ket\uz=\ket{0}^{\otimes{N}}$, the evolved overall state
obeys
\begin{equation}
  e^{-\imath{\cal{H}}t}\ket{\underline\psi}=\alpha\,\ket{\uz}
   +\beta\,\sum_i\bra\ui e^{-\imath{\cal{H}}t}\ket{\uo}\,\ket{\ui}~,
\end{equation}
so that
\begin{equation}
 \rho_i(t)=
 \begin{bmatrix}
  ~1{-}|\beta|^2|u_i(t)|^2 & \alpha\beta^*u_i^*(t) \\
   \alpha^*\beta\,u_i(t) & |\beta|^2|u_i(t)|^2 \\
 \end{bmatrix}~,
\end{equation}
where $u_i(t)$ is the transition amplitude from site 1 to site $i$,
\begin{equation}
  u_i(t) \equiv \bra\ui\,e^{-\imath{\cal{H}}t}\ket{\uo}
  =\sum_{n=1}^N U_{ni}U_{n1}\,e^{-\imath\omega_nt}~.
\label{e.uit}
\end{equation}

The state of the last qubit is described by the density matrix
$\rho_{_N}(t)$, whose degree of similarity with the initial state
$\ket\psi\bra\psi$ can be estimated by the fidelity,
\begin{equation}
 {\cal F}(\psi,t) = \bra{\psi}\rho_{_N}(t)\ket{\psi}~.
\end{equation}
The overall channel quality can be estimated by taking the average of
$F(\psi,t)$ over all possible initial states $\ket\psi$, i.e., over
the Bloch sphere, $|\alpha|^2+|\beta|^2=1$. This yields the average
fidelity ${\cal F}(t)$ in terms of $u_{_N}(t)$, with a dependence on
the phase of the amplitude that one can get rid of by prescribing a
proper local rotation to the last qubit~\cite{BACVV2011}, or, in a
spin-chain context, by applying a suitable magnetic
field~\cite{Bose2003}, in such a way that $u_{_N}(t)$ is real and
positive at the arrival time; with this proviso, setting
\begin{equation}
 u(t)\,{\equiv}\,|u_{_N}(t)|~,
\end{equation}
one has
\begin{equation}
 {\cal F}(t) = \frac13+\frac{[1{+}u(t)]^2}6~.
\label{e.F}
\end{equation}
Hence, the average fidelity is a monotonic function of (the modulus
of) the transition amplitude $u(t)$. On the same footing, it can be
shown that also the entanglement fidelity, that measures the
efficiency of transfer for an entangled state with an external
noninteracting qubit, reads~\cite{BACVV2011}
\begin{equation}
 {\cal F}_{\rm{E}}(t) = \frac{[1{+}u(t)]^2}4~.
\label{e.FE}
\end{equation}
It is clear at this point that $u(t)\in[0,1]$ can be assumed as a
transfer-quality indicator, $u(t)\,{=}\,1$ corresponding to perfect
transfer. For almost perfect transfer, $u\,{=}\,1\,{-}\,\varepsilon$,
one has a reduction of the fidelity to the first order in
$\varepsilon$,
\begin{equation}
 {\cal F} \approx 1 - {\textstyle\frac23}\,{\varepsilon}~,~~~~~
  {\cal F}_{\rm{E}} \approx 1 - {\varepsilon}~.
\end{equation}
In the following we will thus concentrate onto the task of maximizing
the transition amplitude $u(t)$ for the particular model introduced
in the next section.

\subsection{Mirror symmetric case}

If $\bm{A}$ is mirror-symmetric, Eq.~\eqref{e.jn} tells that
$U_{nN}\,{=}\,-e^{\imath\pi{n}}U_{n1}$, so the transition amplitude
we are interested in takes the form
\begin{equation}
 u(t)= \bigg|\sum_{n=1}^N{\cal P}_n
       \,e^{\imath(\pi{n}-\omega_n t)}\bigg| ~,~~~
 {\cal P}_n\equiv U_{n1}^2~,
\label{e.uNt}
\end{equation}
which consists of a sum of time-evolving phases with a weight given
by the mode distribution (or density) ${\cal P}_n$. In the case of
equally spaced eigenfrequencies $\omega_n=\omega_0\,{+}\,v(\pi/N)n$,
where $v$ is a constant, one would have $u(t{=}N/v)\,{=}\,1$, i.e.,
full transmission of the state. As already mentioned in the
Introduction, this case of a `linear dispersion' can be realized in
discrete finite systems~\cite{ChristandlDEL2004} by allowing any
single nearest-neighbor coupling to be properly tuned such that
$A_{i,i+1}\,{\propto}\,\sqrt{i(N{-}i)}$.

\section{The quasi-uniform channel}
\label{s.QUchannel}

\begin{figure}[t]
\includegraphics[width=80mm,angle=0]{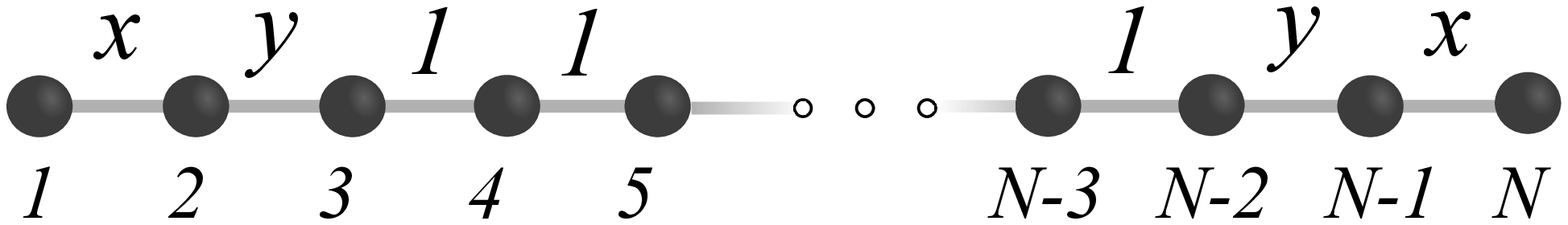}
\caption{Mirror-symmetric quantum channel consisting of $N$ sites with
uniform bonds of unit strength, but for two pairs of weaker bonds at
the extrema, $x$ and $y$. State-transmission is meant to occur
between sites 1 and $N$.}
\label{f.channel}
\end{figure}

We assume from now on that the mirror-symmetric matrix $A=\{A_{ij}\}$
describes the chain of Fig.~\ref{f.channel}, which is uniform except
for the two extremal bonds $A_{12}=A_{N{-}1,N}\equiv{x}$ and
$A_{23}=A_{N{-}2,N{-}1}\equiv{y}$, while all other couplings are
equal and set the energy scale, $A_{i,i{+}1}=1$ for $i=3,...,N{-}3$.
Such quasi-uniform interactions were used in
Refs.~\cite{GiampaoloI2009,GiampaoloI2010} for the purpose of
studying long-distance entanglement in the ground state. Explicitly,
\begin{equation}
A(x,y) =
\begin{pmatrix}
 ~0~ &  x  &      &      &      &     &  &  \\
  x  & ~0~ &  y   &      &      &     &  &  \\
     &  y  & ~0~  & ~1~  &      &     &  &  \\
     &     &  1   & ~0~  &  1   &     &  &   \\
     &     &  &\ddots&\ddots&\ddots&     &     \\
     &     &  &      &  ~1~ & ~0~  &  y  &     \\
     &     &  &      &      &  y   & ~0~ &  x  \\
     &     &  &      &      &      &  x  & ~0~ \\
\end{pmatrix}~.
\label{e.AN}
\end{equation}
It is assumed that $0\,{<}\,x\,{\leq}1$ and $0\,{<}\,y\,{\leq}\,1$.
In Appendix~\ref{a.spectral} the $N$ energy eigenvalues are found in
the form $\omega_n\,{=}\,\cos{k_n}$, where the allowed values
$k\,{=}\,k_n$ are determined by Eqs.~\eqref{e.kn}
and~\eqref{e.shiftk}, and the mode distribution ${\cal P}_k$ is given
by Eq.~\eqref{e.pk}. As the latter is symmetric around $\pi/2$, it is
more convenient to work with the shifted variable
\begin{equation}
 q \equiv \frac\pi{2} - k ~~~~\in~\Big(-\frac{\pi}2,\frac{\pi}2\Big)
\label{e.q}
\end{equation}
whose allowed values are
\begin{equation}
  q_m = \frac{\pi\,m-2\varphi_{q_m}}{N{+}1}
~,~~~
 m=-\frac{N{-}1}2,\dots,\frac{N{-}1}2~,
\label{e.qm}
\end{equation}
where $m\,{\equiv}\,(N{+}1)/2-n$; the corresponding eigenfrequencies
are $\omega_q=\sin{q}$. Understanding that the variable $q$ assumes
the allowed values~\eqref{e.qm}, we can use it in the place of the
eigenvalue index $n={N{+}1}/2-m$. Note that for even $N$ the $m$'s
are half-integer: there is no qualitative difference between the
outcomes for even and odd $N$, the latter case is just more easily
handled numerically. The {\em phase shifts}
\begin{equation}
 \varphi_q = \tan^{-1}\bigg[\frac{y^2\sin{2q}}
            {x^2{-}(2{-}y^2)(1{-}\cos{2q})}\bigg]-2q~
\label{e.shiftq}
\end{equation}
displace the $q$-values from the equally spaced values of the fully
uniform case ($\varphi_q\,{=}\,0$ for $x\,{=}\,y\,{=}\,1$). Note that
the phase shifts do not alter the sequence of the $q_m$'s, which
increase with $m$.

\begin{figure}
\includegraphics[height=82mm,angle=90]{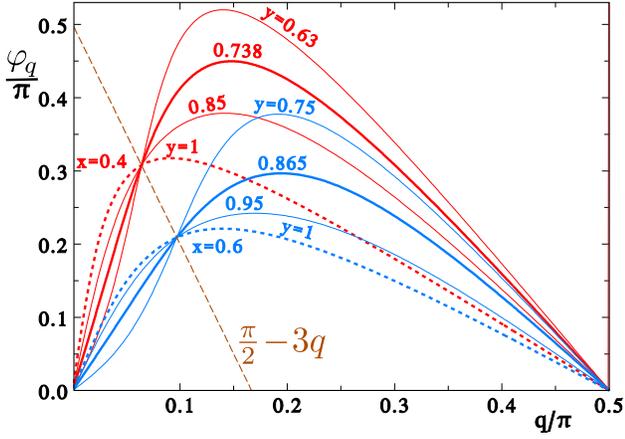}
\caption{(Color online) Shifts $\varphi_q$,
Eq.~\eqref{e.shiftq}, for different values of $x$ and $y$. The dashed
curves correspond to $y\,{=}\,1$. The fixed point
$q_{_{\rm{F}}}\,{=}\,\sin^{-1}\!\frac{x}2$ lies on the dashed
straight line.}
\label{f.shiftq}
\end{figure}

Note that the denominator of Eq.~\eqref{e.shiftq} can vanish and the
argument of $\tan^{-1}$ can jump from $\infty$ to $-\infty$, when
this happens the conventional range of $\tan^{-1}$ has to be extended
above $\pi/2$ rather than jumping from $\pi/2$ to $-\pi/2$. In this
way $\varphi_q$ is a continuous function, shown in
Fig.~\ref{f.shiftq} for positive $q$. The effect of weakening the
bond $y$ is evident: the deformation is such that $\varphi_q$ change
its convexity in an interval $|q|\,{\lesssim}\,q_{_{\rm{F}}}(x)$
whose width is characterized by the fixed point
$q_{_{\rm{F}}}(x)\,{=}\,\sin^{-1}\frac{x}2$. Therefore, the parameter
$y$ can change the separation between the allowed $q$-values, and, in
turn, between the eigenfrequencies affecting the coherence in the
time evolution.

\begin{figure}
\includegraphics[height=82mm,angle=90]{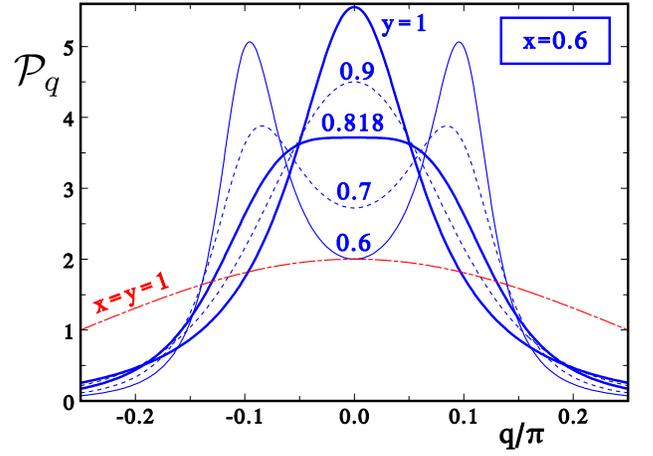}
\caption{(Color online) The normalized large-$N$ mode density
${\cal{P}}_q$, Eq.~\eqref{e.Pq}, for $x\,{=}\,0.6$ and different
values of the second bond coupling $y$. The thicker curves are those
for $y\,{=}\,1$ and for the threshold value $y\,{=}\,Y(x)$ (see text
and Fig.~\ref{f.Yx}). The broad dash-dotted line is the result for
the fully uniform chain, $x\,{=}\,y\,{=}\,1$.}
\label{f.Pq}
\end{figure}

\begin{figure}
\includegraphics[height=82mm,angle=90]{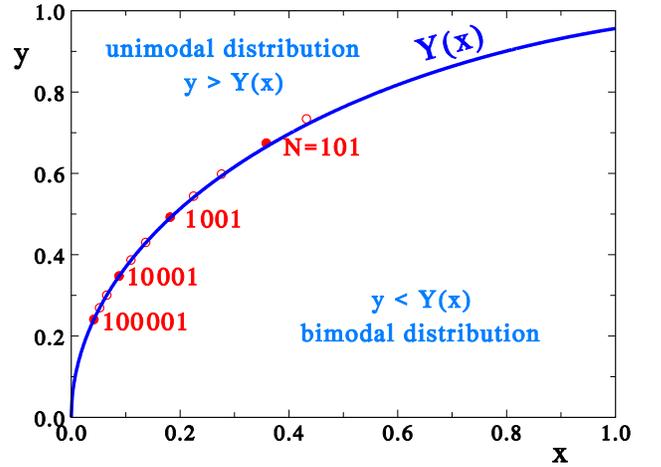}
\caption{(Color online) The function $Y(x)$, Eq.~\eqref{e.y1},
separates the occurrence of unimodal and bimodal mode distributions.
The optimized pairs $(x,y)$, i.e., those which maximize the arrival
amplitude $u(t)$, are also reported for selected values of $N$ (see
Table~\ref{t.opt}).}
\label{f.Yx}
\end{figure}

The mode distribution ${\cal P}_k\,{=}\,U_{k1}^2$, Eq.~\eqref{e.pk},
can be written in terms of $q$ as
\begin{eqnarray}
 {\cal{P}}_q &=& \frac{2}{N{+}1{+}2\varphi_q'} ~\times
\label{e.Pq}\\
 &\times&~\frac{x^2y^2}
   {x^4+(4{-}x^2{-}2y^2)^2\tan^2\!{q}-16\,(1{-}y^2)\sin^2\!{q}}~.~~
\notag
\end{eqnarray}
The term in $\varphi_q'$ is clearly irrelevant for a long chain, even
though it ensures the normalization, $\sum_m{\cal{P}}_{q_m}\,{=}\,1$,
for any finite $N$. Fig.~\ref{f.Pq} describes the typical behavior of
the mode density ${\cal{P}}_q$ when $y$ is varied keeping a fixed
$x\,{<}\,1$. Compared to its counterpart in the case of the uniform
chain, $x\,{=}\,y\,{=}\,1$, it is evidently more structured; by
lowering $y$ its tails get increasingly suppressed and ${\cal{P}}_q$
definitely changes to a bimodal shape with two symmetric maxima: this
occurs when in the denominator the coefficient of $\sin^2\!{q}$
becomes larger than that of $\tan^2\!{q}$, i.e., when $y$ is smaller
than
\begin{equation}
 Y(x) \equiv \sqrt{\sqrt{2}\,x-\frac{x^2}2} ~,
\label{e.y1}
\end{equation}
which is the curve shown in Fig.~\ref{f.Yx}. ${\cal{P}}_q$ is
unimodal for $y\,{\geq}\,Y(x)$ and has its maximum at $q\,{=}\,0$; at
the threshold $y\,{=}\,Y(x)$ the maximum flattens, the deviation
being $\sim{q^4}$, before developing the lateral maxima which are the
more pronounced the smaller $y$. In the limit $y\,{\to}\,0$ the
distribution ${\cal{P}}_q$ tends to two symmetric $\delta$-peaks at
$q=\pm\sin^{-1}\!\frac{x}2$, corresponding to the excitations of the
single dimer with interaction $x$, which is indeed isolated from the
chain when $y\,{=}\,0$.

To represent the behavior of the level spacings, whose uniformity is
crucial for the coherence of transmission, one can define a sort of
`group velocity' by
\begin{equation}
 v_q \equiv \frac{N{+}1}\pi~\omega'_q\,\,\partial_mq
     = \frac{N{+}1}{N{+}1{+}2\varphi_q'}\,\cos{q}~,
\label{e.vq}
\end{equation}
where the `derivative' of Eq.~\eqref{e.qm},
$(N{+}1{+}2\varphi'_q)\,\partial_mq\,{=}\,\pi$, has been used, and
from Eq.~\eqref{e.shiftq}
\begin{equation}
 \varphi_q' = -2 +\frac{2y^2[x^2+2(2{-}x^2{-}y^2)\sin^2\!{q}]}
 {x^4+4[y^4{-}x^2(2{-}y^2)]\sin^2\!{q}+16(1{-}y^2)\sin^4\!{q}}\,.
\end{equation}

\section{Transition amplitude}
\label{s.amplitude}

\subsection{Numerical results}

\begin{figure}
\includegraphics[height=82mm,angle=90]{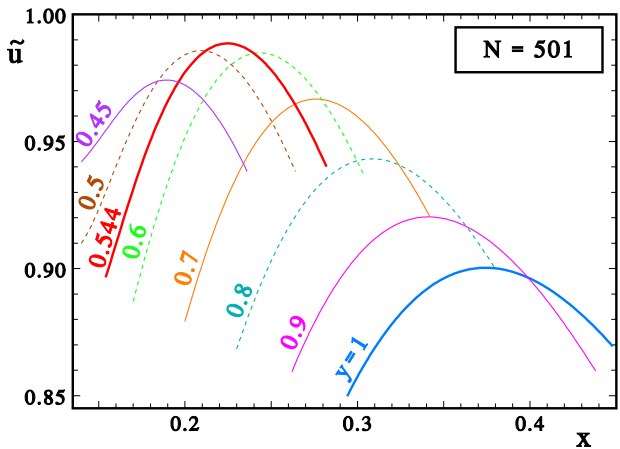}
\caption{(Color online) End-to-end amplitude at arrival
$\tilde{u}(x,y)$ as a function of $x$ for selected values of $y$ and
$N\,{=}\,501$. The thicker curves correspond to $y\,{=}\,1$ and to
the optimized value $y\,{=}\,0.5439$.}
\label{f.u00501}
\end{figure}

\begin{figure}
\includegraphics[width=82mm,angle=0]{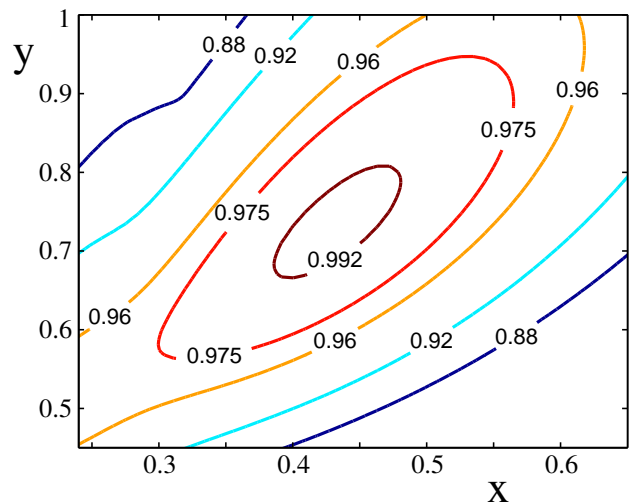}
\caption{(Color online) Contour plot of the average fidelity at arrival
${\cal{F}}(x,y)$ in the $(x,y)$ plane for $N\,{=}\,51$.}
\label{f.f00051c}
\end{figure}

\begin{figure}
\includegraphics[width=82mm,angle=0]{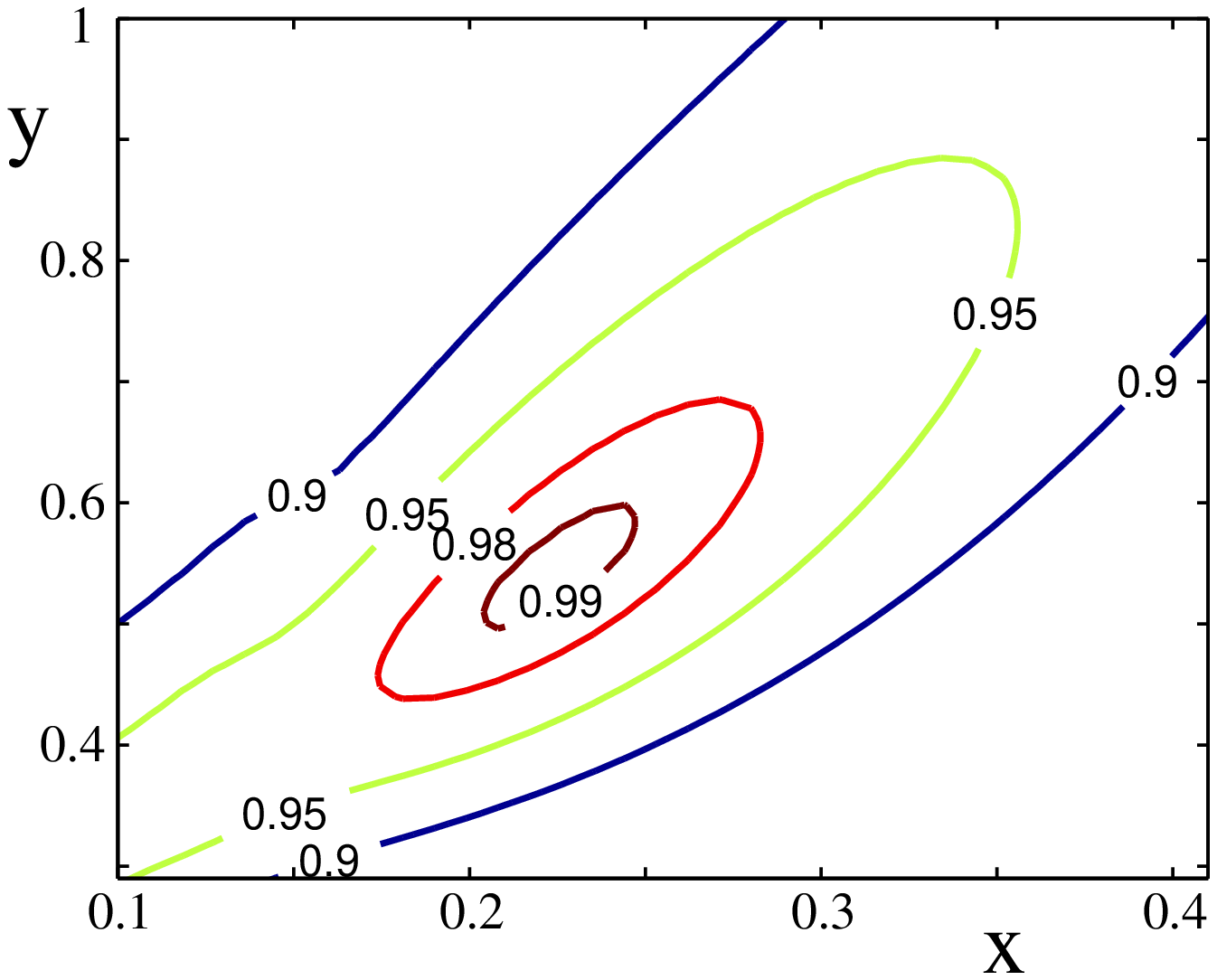}
\caption{(Color online) Contour plot of the average fidelity at arrival
${\cal{F}}(x,y)$ in the $(x,y)$ plane for $N\,{=}\,501$.}
\label{f.f00501c}
\end{figure}

The transition amplitude between the sites $1$ and $N$ at the time
$t$ is given by Eq.~\eqref{e.uNt}. It can conveniently be rewritten
as a sum over the allowed $q$'s~\eqref{e.qm},
\begin{equation}
 u(t) = \Big|\sum_m {\cal{P}}_{q_m}~e^{\imath(\pi m-t\sin q_m)}\Big|~.
\label{e.utq}
\end{equation}
This sum can be evaluated numerically: for any pair $(x,y)$ our code
solves iteratively the coupled Eqs.~\eqref{e.qm} and
~\eqref{e.shiftq} and looks for the value of $t\,{\gtrsim}\,N$, i.e.,
the arrival time, where $u(t)$ attains its largest value, say
$\tilde{u}(x,y)$. Typical results for $\tilde{u}(x,y)$ are reported
as a function of $x$ for selected values of $y$ in
Fig.~\ref{f.u00501}, which refers to a chain of length $N\,{=}\,501$.
One can see that taking $y$ smaller than 1 the amplitude can become
much closer to 1, also for the longest channels, and it is possible
to identify the optimal values $(x^{\rm{opt}},y^{\rm{opt}})$ that
make $\tilde{u}(x,y)$ reach its maximum
$u^{\rm{opt}}\,{\equiv}\,\tilde{u}(x^{\rm{opt}},y^{\rm{opt}})$.
Moreover, the maxima are so broad in the $(x,y)$ plane that a
relatively large mismatch from the optimal values still keeps giving
a large amplitude, and consequently a large average transmission
fidelity, as can be appreciated in the contour plots of
Figs.~\ref{f.f00051c} and~\ref{f.f00501c}.

\begin{table*}
\caption{Optimal pairs $(x^{\rm{opt}},y^{\rm{opt}})$ and the
corresponding amplitude
$u^{\rm{opt}}\,{=}\,\tilde{u}(x^{\rm{opt}},y^{\rm{opt}})$ and average
fidelity ${\cal{F}}^{\rm{opt}}$ for different channel lengths $N$.
Also reported are the optimal values obtained~\cite{BACVV2011}.}
\label{t.opt}
\begin{center}
\begin{tabular}{|r|cccc|ccc|}
\hline
  $N$~~ & $x^{\rm{opt}}$ & $y^{\rm{opt}}$ & $u^{\rm{opt}}$
      & ${\cal{F}}^{\rm{opt}}$ & ~$x^{\rm{opt}}(y{=}1)$~
      & ~$u^{\rm{opt}}$~& ${\cal{F}}^{\rm{opt}}$  \\[1mm]
\hline
      51 & 0.4322 & 0.7338 & 0.99270 & 0.99514 & 0.5542 & 0.9493 & 0.9666 \\
     101 & 0.3584 & 0.6742 & 0.99091 & 0.99395 & 0.4931 & 0.9324 & 0.9557 \\
     251 & 0.2760 & 0.5982 & 0.98932 & 0.99290 & 0.4216 & 0.9127 & 0.9431 \\
     501 & 0.2247 & 0.5439 & 0.98855 & 0.99239 & 0.3742 & 0.9003 & 0.9352 \\
  1\,001 & 0.1818 & 0.4923 & 0.98849 & 0.99235 & 0.3322 & 0.8899 & 0.9286 \\
  2\,501 & 0.1367 & 0.4300 & 0.98765 & 0.99179 & 0.2840 & 0.8791 & 0.9218 \\
  5\,001 & 0.1097 & 0.3869 & 0.98747 & 0.99167 & 0.2523 & 0.8726 & 0.9178 \\
 10\,001 & 0.0878 & 0.3474 & 0.98735 & 0.99159 & 0.2242 & 0.8674 & 0.9145 \\
 25\,001 & 0.0652 & 0.3004 & 0.98726 & 0.99153 & 0.1920 & 0.8621 & 0.9112 \\
 50\,001 & 0.05209& 0.26925& 0.98722 & 0.99151 & 0.1708 & 0.8590 & 0.9093 \\
100\,001 & 0.04150& 0.24072& 0.98720 & 0.99149 & 0.1519 & 0.8565 & 0.9078 \\
\hline
~$N\to\infty$~ &~1.954\,$N^{-1/3}$
               &~1.662\,$N^{-1/6}$~& ~~0.98715~~& ~~0.99146~~
               &~1.030\,$N^{-1/6}$~& ~~0.8469~~ & ~~0.9018~~ \\[0.7mm]
\hline
\end{tabular}
\end{center}
\end{table*}

\begin{figure}
\includegraphics[height=82mm,angle=90]{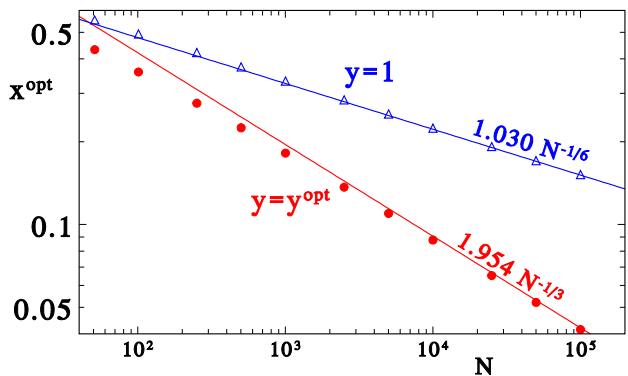}
\caption{(Color online) Optimal values for $x$, from
Table~\ref{t.opt}, reported vs. $N$ both for $y\,{=}\,1$ and for the
optimized $y\,{=}\,y^{\rm{opt}}$. Note that $x\,{\sim}\,N^{-1/6}$ in
the former case, while $x\,{\sim}\,N^{-1/3}$ in the latter.}
\label{f.xyscaling}
\end{figure}

The numerically evaluated optimal data are reported in
Table~\ref{t.opt}, together with those obtained in
Ref.~\cite{BACVV2011} by varying only $x$ with $y\,{=}\,1$. Comparing
with the latter, it is seen that the transfer quality improves in an
extraordinary way: even for a chain of length $N\,{=}\,50\,001$ the
amplitude increases from $0.859$ to $0.987$, i.e., the state-transfer
fidelity ${\cal{F}}$ rises from $0.909$ to $0.991$. The values of
Table~\ref{t.opt} show that the optimized transmission amplitude
decreases more and more weakly for large $N$, so that there could be
a finite asymptotic amplitude attainable for an infinite chain: this
is the case, indeed, as shown in the next subsection. In
Fig.~\ref{f.xyscaling} the optimal values for $x$ are reported for
$y\,{=}\,1$ and for $y\,{=}\,y^{\rm{opt}}$. As shown formally in the
next subsection, it turns out that $x^{\rm{opt}}$ scales as
$N^{-1/6}$ in the former case, while it obeys a new scaling law,
apparently $N^{-1/3}$, in the latter.

\begin{figure}
\includegraphics[height=82mm,angle=0]{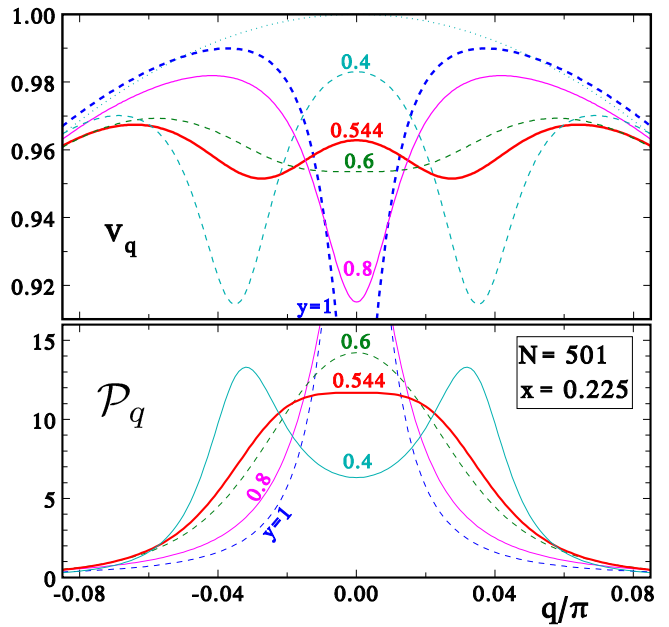}
\caption{(Color online) Group velocity $v_q$ and corresponding mode
density ${\cal{P}}_q$ for $N\,{=}\,501$,
$x\,{=}\,x^{\rm{opt}}\,{=}\,0.225$, and selected values of $y$. The
thicker curves correspond to $y\,{=}\,1$ (dotted) and to the
optimized value $y\,{=}\,y^{\rm{opt}}\,{=}\,0.544$ (solid).}
\label{f.501_vgPq}
\end{figure}

The great improvement in transmission quality deals with the same
argument we gave in Ref.~\cite{BACVV2011}: although a constant group
velocity $v_q$ yields perfect transmission, it is sufficient that
$v_q$ be constant for the $q$-modes excited by the initialization of
the first qubit whose distribution is ${\cal{P}}_q$. The results
illustrated above confirm that the possibility of controlling, by
means of {\em two} parameters ($x$ and $y$), the {\em two} most
relevant features, shape of ${\cal{P}}_q$ and stability of $v_q$,
allows us to obtain an optimal trade-off leading to nearly perfect
transmission.

The weaker second bond $y$ acts indeed on the group velocity, as it
appears in Fig.~\ref{f.501_vgPq}, where the shape of $v_q$ is
reported for the optimized value of $x$, and compared with the
corresponding mode density ${\cal{P}}_q$. With a smaller $y$, the
central dip appearing for $y\,{=}\,1$ is strongly reduced and the
group velocity just shows a small modulation in a rather wide range,
so favoring a coherent dynamics; at the same time ${\cal{P}}_q$
broadens, but not dramatically as it is mainly controlled by $x$ (
see, e.g., Fig.~4 of Ref.~\cite{BACVV2011}); the optimized value
$y\,{=}\,y^{\rm{opt}}$ clearly gives the best result compatible with
the assumed parametrization.

\subsection{Asymptotic behavior for large $N$}

The numerically estimated optimal pairs $(x^{\rm{opt}},y^{\rm{opt}})$
are shown in Fig.~\ref{f.Yx}. They evidently lie almost exactly on
the threshold curve $Y(x)$ separating the unimodal from the
doubly-peaked shape ${\cal{P}}_q$, i.e.,
$y^{\rm{opt}}\,{\simeq}\,Y(x^{\rm{opt}})$. We expect this to hold in
the large-$N$ regime, since the condition $y\,{=}\,Y(x)$ clears the
quadratic terms in the denominator of the density~\eqref{e.Pq}
leaving smaller tails ${\sim}\,q^{-4}$, thus allowing for a more
effective cut of the modes involving the main nonlinear part of
$v_q$. Therefore, instead of considering $y$ as a free parameter, we
fix it to be given by
\begin{equation}
 y\,{\equiv}\,Y(x)=\sqrt{\sqrt{2}\,x-\frac{x^2}2}
                    =2^{1/4}\,x^{1/2}+O(x^{3/2}) ~.
\label{e.yequivy1}
\end{equation}

As $N\to\infty$ the sum~\eqref{e.utq} can be written as an integral,
\begin{equation}
 u_\infty(t) = \lim_{N\to\infty} \int dm~
        {\cal{P}}_{q_m}~e^{\imath(\pi m-t\sin q_m)}~.
\end{equation}
In order to evaluate this asymptotic expression, first note that from
~\eqref{e.qm} one has to set
$\pi\,m\,{=}\,(N{+}1)q_m+2\varphi_{q_m}$, and
$\pi\,dm\,{=}\,(N{+}1{+}2\varphi'_q)\,dq$, so that
\begin{equation}
 u_\infty(t) = \lim_{N\to\infty} \int_{-\frac{\pi}2}^{\frac{\pi}2}
       \frac{dq}\pi~\widetilde{\cal{P}}_q~
        e^{\imath[(N{+}1)q+2\varphi_q-t\sin q]} ~,
\label{e.uinfty}
\end{equation}
where
\begin{equation}
 \widetilde{\cal{P}}_q  = \frac{2\,x^2y^2}
       {x^4+(4{-}x^2{-}2y^2)^2\tan^2\!{q}-16\,(1{-}y^2)\sin^2\!{q}}
\label{e.rhoq}
\end{equation}
is exactly the normalized function reported in Fig.~\ref{f.Pq}. As by
increasing $N$ the optimal distribution gets narrower and narrower,
the denominator can be expanded taking into account the
assumption~\eqref{e.yequivy1}:
\begin{equation}
 \widetilde{\cal{P}}_q \simeq \frac{2^{3/2}x^3}{x^4 + (2q)^4} ~;
\label{e.exprhoq}
\end{equation}
hence, the width of the relevant $q$-region shrinks with $x$. Let us
introduce the scaled variable $\xi\,{=}\,2q/x$, which is of the order
of unity, so that
\begin{equation}
\widetilde{\cal{P}}_q ~dq
  \simeq \frac{\sqrt2\,d\xi}{1+\xi^4} ~.
\end{equation}
As for the phase in Eq.~\eqref{e.uinfty}, the leading term of the
expansion of Eq.~\eqref{e.shiftq} is
\begin{equation}
 \varphi_q \simeq \tan^{-1}\frac{\sqrt2~\xi}{1{-}\xi^2}~,
\end{equation}
and, defining the arrival-time delay $s$ by $t\,{\equiv}\,N{+}1{+}s$,
the remaining terms read
\begin{equation}
 (N{+}1)q-t\sin q = t(q-\sin{q})-s\,q
 \simeq 
  \tau\,\xi^3-\sigma\,\xi ~,
\end{equation}
where
\begin{equation}
 \tau \equiv \frac16\,\Big(\frac{x}2\Big)^3~t
~,~~~~~~~~~
 \sigma \equiv \frac{x}2\,s ~,
\label{e.tausigma}
\end{equation}
are the rescaled counterparts of the arrival time $t\,{\sim}\,N$ and
delay $s\,{\sim}\,N^{1/3}$. Eventually, the asymptotic value of the
amplitude reads
\begin{equation}
 u_\infty(\tau,\sigma) = \frac{\sqrt2}{\pi}\!
   \int\limits_{-\infty}^{\infty}\!\!d\xi~ \frac{\exp\big[
   \imath\big(\tau\,\xi^3{-}\,\sigma\,\xi
   \,{+}2\tan^{-1}\!\frac{\sqrt2~\xi}{1{-}\xi^2}\big)\big]}{1+\xi^4} \,.
\label{e.uinftyxi}
\end{equation}
For a numerical evaluation it is convenient to perform the
substitution $\xi=\tan{z}$, and consider the maximization of
\begin{equation}
 u_\infty = \frac{2\sqrt2}{\pi}\int_0^{\frac\pi{2}}dz~
        \frac{1{+}\tan^2\!{z}}{1{+}\tan^4\!{z}}
        ~\cos\Phi(z)~,
\label{e.uinftyz}
\end{equation}
with
\begin{equation}
   \Phi(z;\tau,\sigma) = \tau\,\tan^3\!{z}-\sigma\,\tan{z}
           +2\tan^{-1}\!{\frac{~\tan{2z}}{\sqrt2}} ~;
\end{equation}
although this phase strongly oscillates for $z$ close to $\pi/2$, the
weighting function makes the numerical convergence easy. The overall
maximum corresponds to $(\tau,\sigma)\,{=}\,(0.15545,3.1645)$, and
amounts to $u_\infty\,{=}\,0.987153$, which is the asymptotic value
reported in Table~\ref{t.opt} together with the asymptotic scaling
resulting from Eq.~\eqref{e.tausigma},
\begin{eqnarray}
 x^{\rm{opt}} &\simeq& 2\Big(\frac{6\tau}N\Big)^{1/3}
                   \simeq 1.954~N^{-1/3} ~,
\notag\\
 y^{\rm{opt}} &=& Y(x^{\rm{opt}}) ~~~\simeq 1.662~N^{-1/6} ~,
\label{e.scaling}
\end{eqnarray}
while the delay scales as
$s\,{=}\,{2\sigma}/{x}\,{\simeq}\,3.239~N^{1/3}$, so that the arrival
time is
\begin{equation}
 t \simeq N+1+3.239~N^{1/3} ~.
\end{equation}

\section{State-transfer dynamics}
\label{s.dynamics}

\begin{figure}
\includegraphics[width=82mm,angle=0]{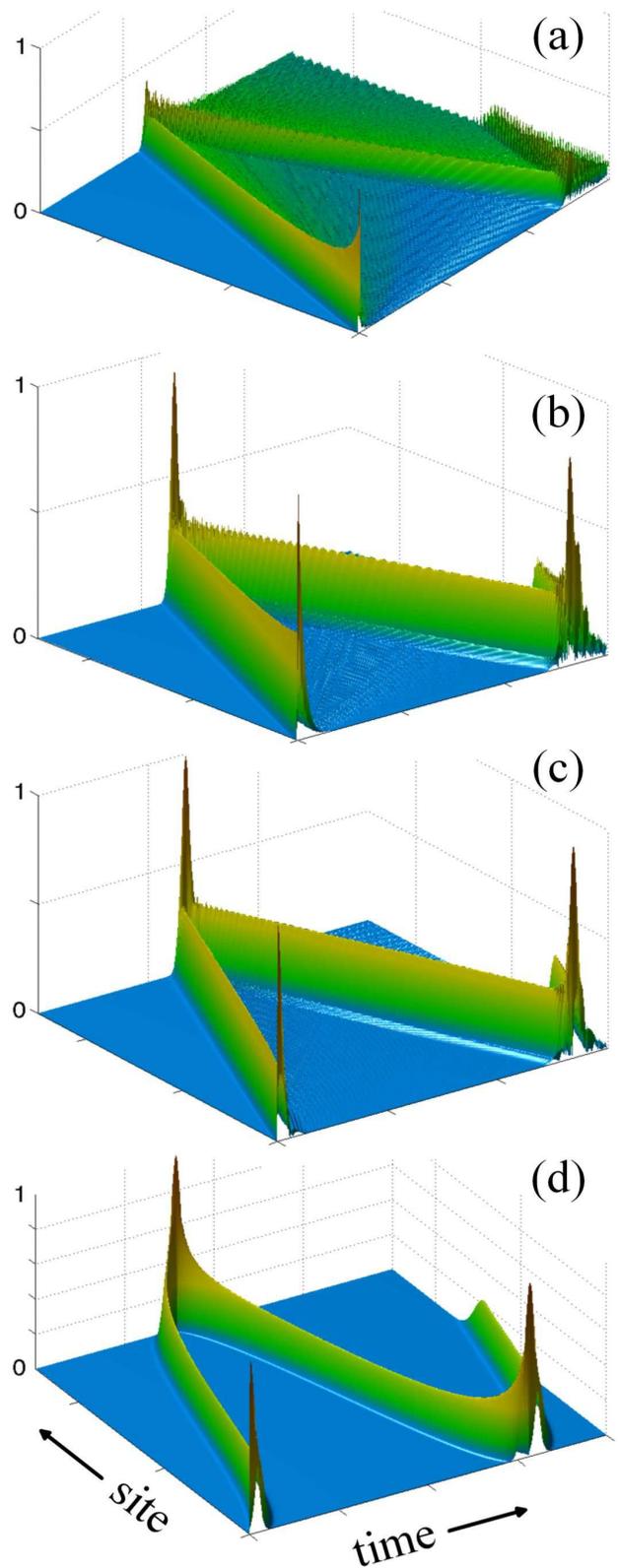}
\caption{(Color online) Space-time perspective views of the
propagating wavepacket $|u_i(t)|$, Eq.~\eqref{e.uit}, for a chain of
length $N\,{=}\,251$, (a) in the fully uniform chain
$x\,{=}\,y\,{=}\,1$, (b) in the case $y\,{=}\,1$ and optimal
$x\,{=}\,0.422$~\cite{BACVV2011}, (c) in the quasi-uniform channel,
Eq.~\eqref{e.AN}, with the optimal $x\,{=}\,0.276$ and
$y\,{=}\,0.598$, and (d) in the perfect-transfer
channel~\cite{ChristandlDEL2004}, Eq.~\eqref{e.christandl}.}
\label{f.u4p}
\end{figure}

\begin{figure}
\includegraphics[width=82mm,angle=0]{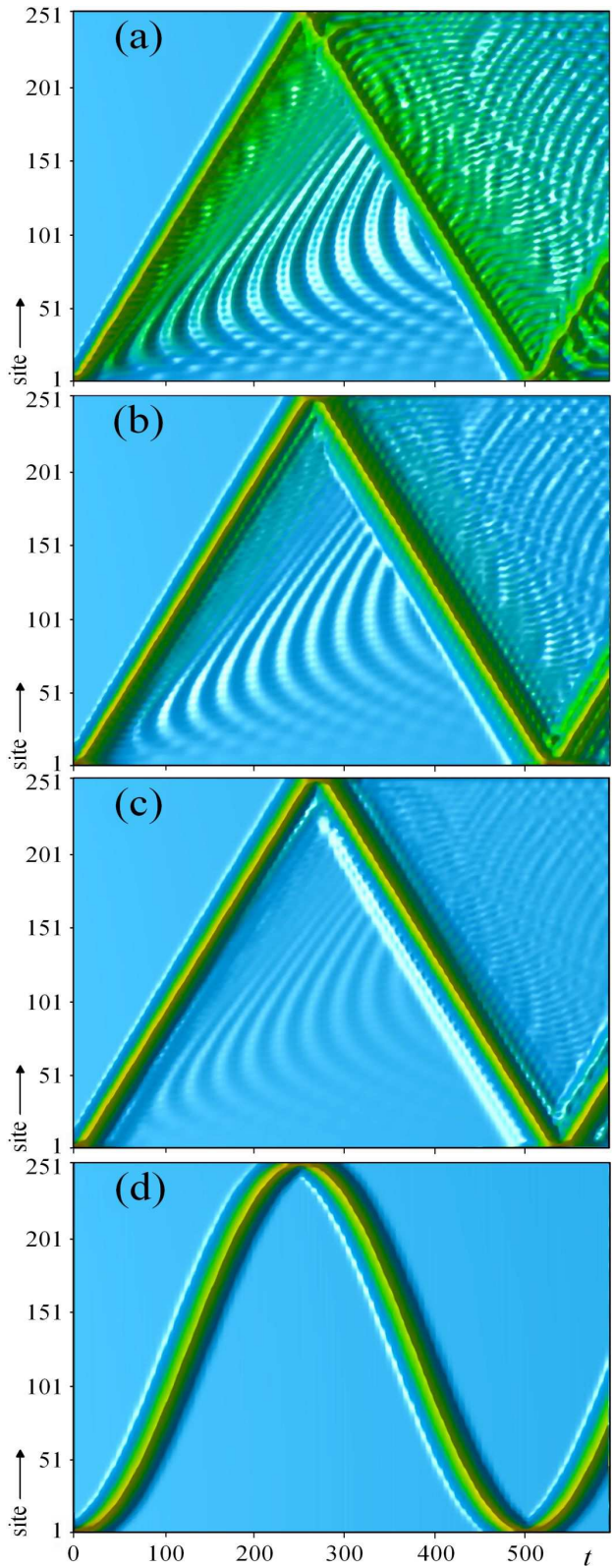}
\caption{(Color online) Space-time contour views of the propagating
wavepacket $|u_i(t)|$, Eq.~\eqref{e.uit} in the same cases of
Fig.~\ref{f.u4p}.}
\label{f.u4c}
\end{figure}

In Section~\ref{s.hoppingmodel} we introduced the instantaneous
transition amplitude from site 1 to any site $i$ of the chain,
Eq.~\eqref{e.uit}. This quantity substantially tells where the
information concerning the initial quantum state sits at any time
$t$. Indeed, it obeys the sum rule
\begin{equation}
  \sum_i |u_i(t)|^2 = \sum_{n=1}^N U_{n1}^2 = 1~,
\label{e.uit2}
\end{equation}
so one can view the state-transfer process as the transmission of a
traveling wavepacket of amplitude $u_i(t)$ which is able to optimally
rebuild most of its content in the single $N$-th site. Looking at the
dynamics of this wavepacket sheds further light onto the ballistic
transfer mechanism and allows for interesting comparisons, based on
the data for $|u_i(t)|$ reported in Figs.~\ref{f.u4p}
and~\ref{f.u4c}.

The first comparison is made in panels~(a), (b) and~(c) of
Figs.~\ref{f.u4p} and~\ref{f.u4c}, and involves: (a) the fully
uniform channel, which displays a very dispersive dynamics and is
indeed inefficient for transmission; (b) the channel with only one
optimized extremal bond~\cite{BACVV2011}, showing an increased
coherence; (c) the further step with two optimized extremal couplings
which improves transmission close to perfection. These features can
be appreciated looking at the height of the arrival-time maxima in
Fig.~\ref{f.u4p}, and at the the rugged features, more evident in
Fig.~\ref{f.u4c}, that represent the amplitude losses due to
dispersion, according to Eq.~\eqref{e.uit2}. Fig.~\ref{f.u4c} makes
also evident the increasing arrival delay from (a) to (b) and to (c),
as a consequence of the slower packet injection and reconstruction
due to the softened endpoint couplings.

Eventually, let us consider the case of perfect state
transfer~\cite{ChristandlDEL2004}, which is obtained by designing all
nearest-neighbor couplings along the chain proportionally to the
height of a semicircle of radius $N{+}1$ drawn over the chain,
\begin{equation}
 A_{i,i+1} = A_{i+1,i} = \frac{\pi}{N{+}1}\sqrt{i(N{-}i)}~;
\label{e.christandl}
\end{equation}
the energy unit being arbitrary, it is chosen here such that the
resulting linear spectrum
\begin{equation}
 \omega_m = \frac{\pi}{N{+}1}~m~,
~,~~~
 m=-\frac{N{-}1}2,\dots,\frac{N{-}1}2~,
\end{equation}
yields the exact arrival time $t\,{=}\,N{+}1$. Notice that,
neglecting terms $\sim{N^{-1}}$, the maximum of the couplings
$\{J_i\}$ is $\pi/2$, their average is $\pi^2/8$, and their
mean-square value is $\pi/3$; this allows for a meaningful comparison
with our model, and the corresponding data are reported in the
panels~(d) of Figs.~\ref{f.u4p} and~\ref{f.u4c}. The modulated
couplings determine a varying velocity of the wavepacket along the
chain, at variance with its constant velocity in the uniform channel
of panels (a-c).

\section{Conclusions}
\label{s.concl}

We have shown that almost perfect ballistic quantum-state transfer
with fidelity larger than $0.99$ can be obtained in an unmodulated
channel of arbitrary length $N$, just by allowing the two endpoint
pairs of nearest-neighbor interactions, $x$ and $y$, to assume
optimal values $(x^{\rm{opt}},y^{\rm{opt}})$. In addition, this
maximum of the transmission quality in the $(x,y)$ plane, as measured
by the average fidelity or equivalently by the transition
amplitude~\eqref{e.uNt}, is so broad that an experimental realization
would not be bound to a fine-tuning of the endpoint couplings, as
Figs.~\ref{f.f00051c} and~\ref{f.f00501c} clearly show.

One might think that the approach presented here looks like the
second step of a sequence that, by allowing further bonds to vary,
would lead to perfect ballistic transmission as found in
Ref.~\cite{ChristandlDEL2004}. However, such deduction does not hold
true: indeed, while the aim of Ref.~\cite{ChristandlDEL2004} is that
of obtaining perfect transfer by letting all normal modes to evolve
coherently, here we look for high-quality transfer by requiring that
only the modes excited by the initialization of the first qubit be
able to evolve coherently. This is done within an effective scheme
ruled by two effects, namely the mode distribution and the frequency
spacings, that can be kept under control by just two parameters,
i.e., the extremal couplings $x$ and $y$. Our approach leads to
extremal couplings $x{\sim}\,N^{-1/3}$ and $y{\sim}\,N^{-1/6}$,
sensibly larger than those required for perfect transfer,
Eq.~\eqref{e.christandl}, which scale as $N^{-1/2}$.

The asymptotic results improve from the uniform chain
($x\,{=}\,y\,{=}\,1$) which gives fidelity
${\cal{F}}_\infty\,{=}\,1/2$, to the single tuned
bond~\cite{BACVV2011} ($y\,{=}\,1$) giving
${\cal{F}}_\infty\,{=}\,0.902$, and eventually to the optimal
asymptotic fidelity evaluated in this paper,
${\cal{F}}_\infty\,{=}\,0.991$. Whether comparable results could be
obtained by other setups is difficult to say: the variants are
numerous and we cannot tell a conclusive word in this respect.

The quasi-uniform channel here considered was previously used in
Refs.~\cite{GiampaoloI2009,GiampaoloI2010} for the different purpose
of exploiting the quasi-long-distance entanglement shared by the
extremal spins in the ground state in order to get efficient
teleportation; in those papers, proposals for realizing the model by
means of coupled cavity arrays and ultracold atoms in 1D optical
lattices are put forward. We expect that such physical realizations
might possibly fit also our setup. To this respect, we remark that
when the presence of disorder in the couplings along the chain is
taken into account, the one-parameter optimal state-transfer scheme
($y=1$) results to be more robust than the perfect-transfer
one~\cite{ZwickASO2012}; we expect such advantage to be preserved in
the two-parameter optimal-transfer scheme.

We finally mention the similarity between the problem of
quantum-state transfer and the subject of continuous-time quantum
walks~\cite{MulkenB005}, where regular space-time structures called
quantum carpets~\cite{Kinzel1995} can emerge from a complex dynamics:
for instance, the revival of the wave-function is analogous, in a
mirror-symmetric context, to state transmission.

\acknowledgements
We acknowledge financial support from the Italian Ministry of
University in the framework of the 2008 PRIN program (contract N.
2008PARRTS 003). TJGA is supported by the European Commission, the
European Social Fund and the Region Calabria through the program POR
Calabria FSE 2007-2013 - Asse IV Capitale Umano-Obiettivo Operativo
M2. LB and PV gratefully thank Dr.~A.~Bayat and, with AC,
Prof.~S.~Bose for useful discussions; TJGA thanks ISC-CNR for the
kind hospitality.

\appendix

\section{Spectral solution}
\label{a.spectral}

\subsection{Characteristic polynomial}

The characteristic polynomial
$\chi_{_N}(\lambda;x,y)\equiv\det[\lambda-A(x,y)]$  associated to the
matrix~\eqref{e.AN} is
\begin{equation}
 \chi_{_N}(\lambda)=
\begin{vmatrix}
 \lambda&  -x   &       &      &       &       &       \\
    -x  &\lambda&  -y   &      &       &       &       \\
        &  -y   &\lambda& -1   &       &       &       \\
        &      & -1 &\lambda& -1   &       &       &       \\
        &      & &\ddots &\ddots&\ddots &       &       \\
        &      & &       &  -1  &\lambda&  -y   &       \\
        &      & &       &      &  -y   &\lambda&  -x   \\
        &      & &       &      &       &  -x   &\lambda\\
\end{vmatrix}_N
  ~.
\label{e.chiN}
\end{equation}

Expanding $\chi_{_N}(\lambda;x,y)$ in the last column one finds
\begin{equation}
 \chi_{_N} = (\lambda^2-x^2)~\xi_{_{N-2}}-\lambda y^2~\xi_{_{N-3}}~,
\label{e.chixi}
\end{equation}
where $\xi_{_N}$ is the characteristic polynomial of the associated
matrix with only one nonuniform endpoint,
\begin{equation}
 \xi_{_N}(\lambda) =
\begin{vmatrix}
 ~\lambda~ & -x      &         &       &         &         \\
  -x       & \lambda &  -y     &       &         &         \\
           &  -y     & \lambda & -1    &         &         \\
           &         & -1 &\lambda & -1    &         &         \\
           &        & & \ddots  &\ddots & \ddots  &         \\
           &        & &         & -1    & \lambda & -1      \\
           &        & &         &       & -1      & \lambda \\
\end{vmatrix}_{N}~,
\end{equation}
which can be expanded in the very same manner getting the analog of
Eq.~\eqref{e.chixi},
\begin{equation}
 \xi_{_N} = (\lambda^2-x^2)~\eta_{_{N-2}}-\lambda y^2~\eta_{_{N-3}}~,
\label{e.xieta}
\end{equation}
in terms of the characteristic polynomial
$\eta_{_N}(\lambda)\equiv\chi_{_N}(\lambda;1,1)$ of the fully uniform
matrix $A_{_N}(1,1)$. For $\eta_{_N}$ one has the recursion relation
\begin{equation}
 \eta_{_N}= \lambda\,\eta_{_{N-1}}-\,\eta_{_{N-2}}~;
 \label{e.etaeta}
\end{equation}
together with the conditions $\eta_{_0}=1$ and $\eta_{_1}=\lambda$,
it can be solved in terms of Chebyshev polynomials of the second
kind,
\begin{equation}
  \eta_{_N} = \frac{\sin(N{+}1)k}{\sin{k}}
~,~~~~
 \lambda \equiv 2\cos k ~.
\label{e.chebyshev}
\end{equation}
Hence, in the uniform case the $N$ solutions of the secular equation
$\chi_{_N}(\lambda;1,1)=\eta_{_N}(\lambda)=0$
correspond~\cite{LiebSM1961} to the following discrete values of $k$:
\begin{equation}
 k = \frac{\pi\,n}{N{+}1}~,~~~n=1, \dots, N~.
\label{e.k_x1y1}
\end{equation}

To extract the dependence on $x$ and $y$ in Eq.~\eqref{e.chixi} one
first uses Eq.~\eqref{e.xieta},
\begin{eqnarray}
 \chi_{_N}(\lambda)
 &=& (\lambda^2{-}x^2)\big[(\lambda^2{-}x^2)\eta_{_{N-4}}
                          -\lambda y^2\eta_{_{N-5}}\big]
\notag\\
   &&~~~-\lambda y^2\big[(\lambda^2{-}x^2)\eta_{_{N-5}}
                    -\lambda y^2\eta_{_{N-6}}\big]~,
\end{eqnarray}
and then Eq.~\eqref{e.chebyshev} in the form
$\eta_{_N}\sin{k}=\Im\big\{e^{\imath(N{+}1)k}\big\}$,
\begin{eqnarray}
 \chi_{_N}(\lambda)\,\sin{k}
 &=& \Im\big\{\big(\lambda^2{-}x^2{-}\lambda y^2e^{-\imath k}\big)^2
 e^{\imath(N{-}3)\imath}\big\}
\notag\\
 &=& \Im\big\{ u_k^2\,e^{\imath(N{+}1)k} \big\}~,
\label{e.chik}
\end{eqnarray}
where
\begin{eqnarray}
 u_k &\equiv& (\lambda^2{-}x^2{-}\lambda y^2e^{-\imath k})\,e^{-2\imath k}
\notag \\
 &=& 1+(2{-}x^2{-}y^2)e^{-2\imath k}+(1{-}y^2)e^{-4\imath k}~.~~
\label{e.uk}
\end{eqnarray}

\subsection{Shifts of the eigenvalues}

Eq.~\eqref{e.chik} leads to a compact expression of the secular
equation, $\Im\big\{e^{\imath(N{+}1)k}\,u_k^2\big\}=0$; in the fully
uniform case, when both $x,y\to{1}$, it has the solutions given by
Eq.~\eqref{e.k_x1y1}. Setting
\begin{equation}
 u_k \equiv |u_k|\,e^{-\imath\varphi_k}~,
\label{e.ukpolar}
\end{equation}
the secular equation reads
$\Im\{e^{\imath[(N{+}1)k-2\varphi_k]}\}\,{=}\,0$, and the eigenvalues
$\lambda_n\,{=}\,2\cos{k_n}$ can be expressed as deviations from the
uniform-case values~\eqref{e.k_x1y1} due to the {\em phase shifts}~
$\varphi_k$,
\begin{equation}
  k_n = \frac{\pi\,n+2\varphi_{k_n}}{N{+}1}  ~,~~~n=1, \dots, N~.
\label{e.kn}
\end{equation}
Keeping in mind that the variable $k$ actually assumes the $N$
discrete values $\{k_n\}$, one can unambiguously use the index $k$ in
the place of $n$. An explicit expression of the phase shifts follows
immediately from Eq.~\eqref{e.uk}, by separating the real and
imaginary parts of $u_ke^{2\imath{k}}$,
\begin{equation}
 \varphi_k = 2k-\tan^{-1}\bigg[\frac{y^2\sin{2k}}
            {(2{-}x^2{-}y^2)+(2{-}y^2)\cos{2k}}\bigg]~.~
\label{e.shiftk}
\end{equation}

\subsection{Mode distribution}

To evaluate the transition amplitude~\eqref{e.uNt} one needs the
square components of the first column of the orthogonal matrix $U$
defined by Eq.~\eqref{e.Uni}. A nice formula derived in
Ref.~\cite{Parlett1998} (Corollary 7.9.1) comes to help: with the
formalism used here it reads
\begin{equation}
 {\cal P}_n \equiv U_{n1}^2 = \frac{\chi_{_{2:N}}(\lambda_n)}
 {\partial_\lambda\chi_{_N}(\lambda_n)}~,
\end{equation}
where $\chi_{_{2:N}}(\lambda)$ is the first minor of the
determinant~\eqref{e.chiN}. It is equivalent to use the variable
$k\,{=}\,\cos^{-1}\frac\lambda{2}\in(0,\pi)$ and write
\begin{equation}
 {\cal P}_k \equiv U_{k1}^2 =
 -2\sin{k}~\frac{\chi_{_{2:N}}(k)}{\partial_k\chi_{_N}(k)}~,
\label{e.Uk1}
\end{equation}
with $k$ definitely taking the allowed values~\eqref{e.kn}, i.e.,
$\chi_{_N}(k)=0$. As
$\chi_{_{2:N}}(\lambda)=\lambda\,\xi_{_{N-2}}{-}y^2\xi_{_{N-3}}$,
with calculations similar to those which lead to Eq.~\eqref{e.chik}
it is found that
\begin{equation}
 \chi_{_{2:N}}(k)\,\sin{k} = \Im\big\{e^{\imath Nk} u_kv_k\big\} ~.
\end{equation}
with
\begin{equation}
 v_k = 1+ (1{-}y^2) e^{-2\imath k} ~.
\label{e.vk}
\end{equation}
By deriving Eq.~\eqref{e.chik} with respect to $k$ one has
\begin{eqnarray}
 \sin{k}~\partial_k\chi_{_N}(k)
 &=& \Im\big\{e^{\imath(N{+}1)k}~[\imath(N{+}1)u_k^2+2u_ku'_k]\big\}
\notag\\
 &=& (N{+}1)\,\Re\big\{e^{\imath(N{+}1)k}~u_k^2 \big\}+
\notag\\
    &&~~~~~~~  +2\,\Im\big\{e^{\imath(N{+}1)k}~u_ku'_k\big\}~;
\end{eqnarray}
the argument of $\Re$ is real by the secular equation so, using
$u_k'/u_k=\partial_k\ln{u_k}=\partial_k\ln|u_k| -i\varphi_k'$,
\begin{eqnarray}
 \sin{k}~\partial_k\chi_{_N}(k)
 &=& \Big[N{+}1+2\,\Im\Big\{\frac{u'_k}{u_k}\Big\}\Big]
 \,e^{\imath(N{+}1)k}\,u_k^2
\notag\\
 &=& (N{+}1-2\varphi_k')\,e^{\imath(N{+}1)k}\,u_k^2~.
\label{e.dkchi}
\end{eqnarray}
Eq.~\eqref{e.Uk1} becomes
\begin{eqnarray}
 {\cal P}_k &=& -\frac{2\sin{k}}{N{+}1{-}2\varphi_k'}
    ~\frac{\Im\{e^{\imath Nk}u_kv_k\}}{e^{\imath(N{+}1)k}u_k^2}
\notag\\
  &=& \frac{2\sin{k}}{N{+}1{-}2\varphi_k'}
  ~\frac{\Im\{e^{\imath k}u_kv_k^*\}}{|u_k|^2} ~.
\end{eqnarray}
By means of Eqs.~\eqref{e.uk} and~\eqref{e.vk} one can express
$u_k=v_k(1{+}e^{-2\imath{k}})-x^2e^{-2\imath{k}}$, and a simple
expression of the numerator follows,
$\Im\{e^{\imath{k}}u_kv_k^*\}=x^2y^2\,\sin{k}$, finally yielding
\begin{equation}
 {\cal P}_k = \frac{2x^2y^2}{N{+}1{-}2\varphi_k'}~
 \frac{\sin^2\!{k}}{|u_k|^2}~.
\label{e.pk1}
\end{equation}
A manageable expression for $|u_k|^2$ arises by working out
Eq.~\eqref{e.uk},
\begin{equation}
 e^{2\imath k}u_k = -x^2{+}2(2{-}y^2)\cos^2\!{k} +2\imath\,y^2\sin{k}\,\cos{k}~,
\end{equation}
and taking the square modulus, after some algebra the outcome is
\begin{eqnarray}
 {\cal P}_k &=& \frac{2x^2y^2}{N{+}1{-}2\varphi_k'}~\times
\label{e.pk}\\
 &\times&~\frac{1}
   {x^4 + (4{-}x^2{-}2y^2)^2\cot^2\!{k}-16\,(1{-}y^2)\cos^2\!{k}}~.
\notag
\end{eqnarray}

An explicit expression for $\varphi_k'$ can be found going back to
Eq.~\eqref{e.dkchi}
\begin{equation}
 \varphi_k' = -\Im\bigg\{\frac{u_k'}{u_k}\bigg\}
            = -\frac{\Im\{u_k^*u_k'\}}{|u_k|^2}~,
\end{equation}
where, from Eq.~\eqref{e.uk},
\begin{equation}
 u_k' =  -2\imath\big[u_k-1-(1{-}y^2)e^{-4\imath k}\big] ~,
\end{equation}
and some further calculation gives
\begin{eqnarray}
 |u_k|^2\varphi_k'
  &=& 2\Re\big\{u_k^*\,\big[u_k-1-(1{-}y^2)e^{-4\imath k}\big]\big\}
\notag\\
  &=& 2|u_k|^2-2y^2[x^2+2(2{-}x^2{-}y^2)\cos^2\!{k}]~.~~~~~
\label{e.dphik}
\end{eqnarray}
Eventually, the mode density can be made fully explicit starting
again from Eq.~\eqref{e.pk1},
\begin{widetext}
\begin{equation}
 {\cal P}_k = \frac{2x^2y^2\sin^2\!{k}}
   {(N{-}3)\{[x^2-2(2{-}y^2)\cos^2\!{k}]^2+4y^4\cos^2\!{k}\sin^2\!{k}\}
     +4y^2[x^2+2(2{-}x^2{-}y^2)\cos^2\!{k}]}~.
\label{e.pkex}
\end{equation}
\end{widetext}

\end{document}